 \documentclass[12pt]{article}

\usepackage{amsmath}
\usepackage{amsfonts}
\usepackage{amssymb}
\usepackage{graphicx}
\usepackage{bm,bbm}
\usepackage{epsfig}
\usepackage{latexsym}
\usepackage{url}
\usepackage{cite}

\def\ep{\varepsilon}

\def\tr{\hbox{tr}}
\def\calm{{\mathcal{M}}}
\def\caln{{\mathcal{N}}}
\def\calb{{\mathcal{B}}}

\def\calh{{\mathcal{H}}}
\def\calk{{\mathcal{K}}}

\def\calc{{\mathcal{C}}}

\def\bra{\langle}
\def\ket{\rangle}

\def\pmx{\begin{pmatrix}}
\def\emx{\end{pmatrix}}

\def\R{\mathbb{R}}
\def\C{\mathbb{C}}
\def\dg{\dagger}

\def\vol{{\rm vol}}
\def\vrad{{\rm vrad}}

\newcommand{\E}{\mathbb{E}}

\newcommand{\N}{\mathbb{N}}

\newcommand{\cB}{\mathcal{B}}

\newcommand{\cH}{\mathcal{H}}

\newcommand{\cM}{\mathcal{M}}

\begin{document}

\title
%
{How often is a random quantum state $k$-entangled?}

\author
{Stanis{\l}aw J. Szarek$^{1,2}$, Elisabeth Werner$^{1,3}$,
and  Karol \.Zyczkowski$^{4,5}$
\smallskip \\
$^1${\small Universit{\'e} Paris Pierre et Marie Curie--Paris 6, France} \\
$^2${\small Case Western Reserve University, Cleveland, Ohio 44106-7058, USA} \\
$^3${\small Universit{\'e} de  Lille 1, 
France} \\
$^4${\small Institute of Physics, Jagiellonian University, 
ul Reymonta 4, 30-059 Krak{\'o}w, Poland}\\
$^5${\small Center for Theoretical Physics, Polish Academy of Sciences
}}

  \date{{\small October 7, 2010}}

\maketitle

\begin{abstract}
The set of trace preserving, positive maps acting on 
density matrices of size $d$ forms a convex body.
We investigate its nested subsets consisting 
of $k$-positive maps, where $k=2,\dots, d$.
Working with the measure induced by the Hilbert-Schmidt
distance we derive asymptotically tight 
bounds for the volumes of these sets.
Our results strongly suggest that the inner set of $(k+1)$--positive maps 
forms a small fraction of the outer set of $k$--positive maps.
These results are related to analogous bounds for the
relative volume of the sets of $k$--entangled states
describing a bipartite $d\times d$ system.
\end{abstract}

\vskip1cm PACS: 03.65.Aa, 03.67.Mn, 02.40.Ft

\newpage

\section{Introduction}

The structure of the set of entangled quantum states 
is  a subject of  vivid scientific interest
in view of  possible applications 
in the theory of quantum information processing.
However, even in the simplest case of systems composed 
of two subsystems only, several basic problems related to the phenomenon 
of quantum entanglement remain unsolved.
For instance, sufficient and necessary conditions for separability 
of an arbitrary quantum state are established only in the case of
four and six dimensional Hilbert spaces ${\C}^4={\C}^2 \otimes {\C}^2$
and  ${\C}^6={\C}^2 \otimes {\C}^3$. 

In higher dimensions several conditions are known 
which imply the property of quantum entanglement \cite{HHH09}.
Unfortunately these tools are not universal and even in the case of
a $3 \times 3$ system there exist quantum states,
the entanglement of which cannot be diagnosed with the general 
criteria currently available.
The structure of  the set of separable states
is thus not well understood, and  its geometry is still  a subject of 
recent studies \cite{BZ06,SBZ06,BS10}.

The complexity of the set of quantum states for a $d \times d$ system
increases quickly with the dimension.
For  $d\ge 3$ it is useful to distinguish 
different degrees of quantum entanglement.
A state $\rho$, represented by a Hermitian, positive semi-definite matrix
with  unit trace (density matrix or density operator),  
is called {\sl separable}, or $1$--{\sl entangled}, 
if it belongs to the convex hull of the set of product states \cite{werner}.  
More  generally, for $1 \leq k \leq d$, one introduces the set of $k$-entangled states,  
the states with  Schmidt rank not larger than $k$ \cite{TH00} (see section \ref{sec:nota} for a precise definition).
Any $k$--entangled state  belongs by definition to the
larger set of $(k+1)$--entangled states and,  in this convention,  the set of
$d$--entangled states coincides with the set of all 
states of a bi--partite,  $d \times d$ system.

For large $d$, the set of separable mixed states of a bipartite $d \times d$ system
is known to cover only a small fraction of the entire body 
of mixed states of size $d^2$. Asymptotically sharp 
estimates for these ratios are known 
and  bounds for the radius of the maximal ball inscribed 
into the set of separable mixed states were obtained, both in 
the bipartite and in the multipartite case  
\cite{ZHSL98,BCJLPS99,GB, GB2, s4,GB3,AS,Hi07}.
The ratio between these volumes depends additionally
 on the measure used \cite{Zy99,Sl04,Ye1,Ye2}.

The structure of the set of states of a composite, bi--partite system
is closely related to properties of the set of linear
maps  that send the convex body of normalized
 mixed states of a mono-partite system into itself.
A map is called {\sl positive} 
if any positive (semi-definite) matrix
is mapped into a positive matrix. 
If $k \geq 1$, a map $\Phi$ 
 is called {\sl $k$--positive}
if the extended  map $\Phi \otimes \ \mathcal {I}_k$ is positive 
(here $\mathcal {I}_k$ is the identity map on ${\cal M}_{k} $, 
the space of $k\times k$ matrices).
A map $\Phi$ is {\sl completely positive}  (CP)
if the extended  map is positive for all $k \in \N$.
Note that if a map $\Phi: {\cal M}_{d} \rightarrow {\cal M}_{d}$ 
is $d$-positive, it is also completely positive \cite{Choi}, 
and so only the range $1\leq k \leq d$ is of interest.
In general, the characterization of a set of
$k$--positive maps is not easy \cite{CK08,SSZ}, 
and the geometry of the set of maps
acting on ${\cal M}_{d}$ is nontrivial even
in the simplest case of $d=2$ \cite{RSW02,BZ06}.

The correspondence between the sets of quantum maps
and quantum states can be made precise due to 
the Choi--Jamio{\l}kowski isomorphism:
the set of trace preserving, completely positive maps
acting on ${\cal M}_{d}$ 
is isomorphic with the set of these bi--partite 
states from ${\cal M}_{d^2}$, for which 
the partial trace over one (say, the second) subsystem 
is proportional to the identity matrix \cite{ZB04,AKMS06}.
Furthermore, the (larger) set of $k$--positive 
  maps is  the dual  of the set
of $k$--superpositive maps (see section \ref{sec:duality} for details and fine points).
These maps, also called $k$--entanglement breaking channels,
correspond by the Choi--Jamio{\l}kowski isomorphism
to $k$--entangled states of a bipartite system \cite{RA07,SSZ}.
Thus investigating relations between sets of 
maps of different degree of positivity one can 
establish properties of the subsets of ${\cal M}_{d^2}$
characterized by different classes of 
quantum entanglement \cite{SWZ,SSZ}, and {\em vice versa}.

One of the main objects of this work is to derive bounds for the volume radius
of the convex body 
of trace preserving, $k$--positive
maps acting on a $d$--dimensional system.
We are working with respect to the Hilbert-Schmidt measure, induced by the
Euclidean (Hilbert--Schmidt, or Frobenius) distance in the space of quantum maps.
These results allow us to estimate the ratio of the volumes
of the different sets. In large dimension,  one finds that the property of 
``additional degrees of positivity" is very rare.
While our methods allow only to compare the set of  $k$--positive maps 
with that of $ak$--positive maps, where $a>1$ in a universal constant 
(independent of $k,d$, but possibly not-so-small), the results obtained 
strongly suggest that the set of $(k+1)$--positive maps 
occupies only a small part of the larger set of  $k$--positive maps, 
with the trend particularly pronounced for small $k$'s.

To arrive at these conclusions,  we study first the volumes of the nested sets of 
$k$--entangled states of a $d\times d$ bi-partite system,
with $k=1,2,\dots, d$. 
Ratio of these volumes can be used to estimate 
the probabity that a random quantum density matrix of a bipartite system 
distributed according to the Hilbert-Schmidt measure \cite{SZ04}
is $k$--entangled.
Of special importance is the case $k=2$,
as  knowledge about the set of $2$--entangled states is crucial
for problems related to the distillation of quantum entanglement \cite{Cl05}.
It would be of substantial interest to rigorously show that, 
for large dimension $d$, the set of $2$--entangled states is 
small in comparison to the set of $3$-entangled states, but 
large with respect to the set of separable states.

\smallskip
This work is organized as follows. In the next section
we introduce some necessary definitions 
involving the sets of trace preserving $k$--positive maps 
and the set of normalized, $k$--entangled states of a 
bi--partite system. The duality relations between convex cones 
and tools used to estimate volumes of dual sets are discussed in section \ref{sec:duality}.
The main results of this paper are obtained in section \ref{sec:volrad4}, where
we derive bounds for the volume radius of the set 
of normalized $k$--entangled states and related parameters.
A summary of the results obtained and their discussion
is presented in the final section \ref{sec:conclusion}.

\section{Trace preserving $k$-positive maps  
        and normalized $k$-entangled states: concepts and notation}
\label{sec:nota}

In this section we recall necessary definitions and introduce notation
used throughout the paper.
\par
If $\cH$ is a Hilbert space, we will denote by $|\cdot |$ its norm and by $\cB(\cH)$ 
the space of bounded linear operators on $\cH$. 
Most often we will have  $\cH = \C^d$ for some $d \in \N$, then operators are 
just matrices and we will write $\cM_d$  for $\cB(\C^d)$.  
We will generally use the Dirac bra-ket notation, but in some contexts we will 
employ the (more common in the functional analysis literature) 
symbols $\bra \cdot , \cdot\ket$ and  $( \cdot , \cdot )$ for the scalar product.

Transformations that are discrete in time can be described
by linear {\sl quantum maps}, or {\em super-operators}, $\Phi: {\calm}_d \to {\calm}_d$ 
(or, more generally, $\Phi: {\calm}_{d_1} \to {\calm}_{d_2}$).
A map is called {\sl positive} (or {\em positivity-preserving})
if every positive (semi-definite) operator
is mapped into a positive operator. 

Let $1 \leq k \leq d$.
A map $\Phi: {\calm}_d \to {\calm}_d$ is called {\sl k-positive}  
if the extended  map $\Phi \otimes \ \mathcal {I}_k  : {\calm}_d \otimes  {\calm}_k\ \to {\calm}_d \otimes {\calm}_k  $
is positive, where $\mathcal{I}_k$ is the identity map on 
${\mathcal M}_k$. 
The set of $k$-positive  maps on ${\calm}_d$ is a  convex cone and will be denoted by 
$\mathcal{P}_k(\mathcal{M}_d)$.

\par
A map $\Phi$ is $k$-positive  iff the  Choi matrix $C_\Phi= \sum _{i,j=1}^d E_{ij}\otimes  \Phi (E_{ij})$ is {\sl $k$-block positive},  i.e. if 
\begin{equation}\label{k-pos}
\Big\langle C_{\Phi} \Big(\sum_{i=1}^k u_i \otimes v_i \Big), \sum_{j=1}^k u_j \otimes v_j \Big\rangle \geq 0
\end{equation}
 for all $u_i,  v_j\in \mathbb{C}^d$, $1 \leq i,j \leq d$ (see, e.g., \cite{SSZ}).
Thus we  can identify the cone $\mathcal{P}_k(\mathcal{M}_d)$ via the Jamio{\l}kowski--Choi isomorphism $\Phi \rightarrow C_\Phi$ with the cone of $k$-block positive operators on $\mathbb{C}^d\otimes \mathbb{C}^d$,
$$
\mathcal{P}_k(\mathcal{M}_d) 
\sim \mathcal{BP}_k(\mathbb{C}^d\otimes \mathbb{C}^d).
$$
\par
As is apparent from condition (\ref{k-pos}),
 the cone $\mathcal{BP}_k(\mathbb{C}^d\otimes \mathbb{C}^d)$ 
is dual to the cone of $k$-entangled operators 
on $\mathbb{C}^d \otimes \mathbb{C}^d$, i.e., to 
\begin{equation} \label{kEnt}
{\rm Ent}_k(\mathbb{C}^d\otimes \mathbb{C}^d)=  \text{conv}\left(\Big\{ |\xi \ket \bra \xi | \, : \, 
\xi = \sum_{j=1}^k  u_j \otimes v_j, \; u_j, v_j  \in \mathbb{C}^d \  {\rm for } \  j=1, \ldots , k \Big\}\right)
\end{equation}
Vectors of the form 
$\xi = \sum_{j=1}^k  u_j \otimes v_j$
will be called  $k$-entangled.
Observe that the special case of $k=1$ 
coincides with the definition of separable (product) matrices or vectors.

A map $\Phi: {\calm}_d \to {\calm}_d$ is said to be {\sl k-super positive}  if its Choi matrix $C_\Phi$ is $k$-entangled. 
This turns out to be equivalent to the existence of a Kraus representation 
$\Phi(\rho) = \sum_i A_{i}^{\dg} \rho A_{i}$, 
of  $\Phi$ such that all the operators $A_i$ 
are of rank smaller than or equal  to $k$ \cite{SSZ}.
 We denote the convex cone  of
$k$-superpositive maps on $ {\calm}_d$ by $\mathcal{SP}_k(\mathcal{M}_d)$.  
As before, we have the identification
$$
\mathcal{SP}_k(\mathcal{M}_d) \ \sim \ 
{\rm Ent}_k(\mathbb{C}^d\otimes \mathbb{C}^d)
$$
and, in the appropriate sense, the cones of maps $\mathcal{P}_k(\mathcal{M}_d)$ and $\mathcal{SP}_k(\mathcal{M}_d)$ are dual to each other.  Note that 
$\mathcal{SP}_d (\mathcal{M}_d)=\mathcal{P}_d (\mathcal{M}_d)= \mathcal{CP}(\mathcal{M}_d)$, 
the convex cone  of   completely positive maps, while for an arbitrary $k \in \N$ we have 
$\mathcal{SP}_k (\mathcal{M}_d)\subset  \mathcal{CP}(\mathcal{M}_d) \subset \mathcal{P}_k (\mathcal{M}_d)$.

\vskip 0.3cm
\begin{figure}[htbp]
  \centering
 \includegraphics[width=.99\textwidth{}]{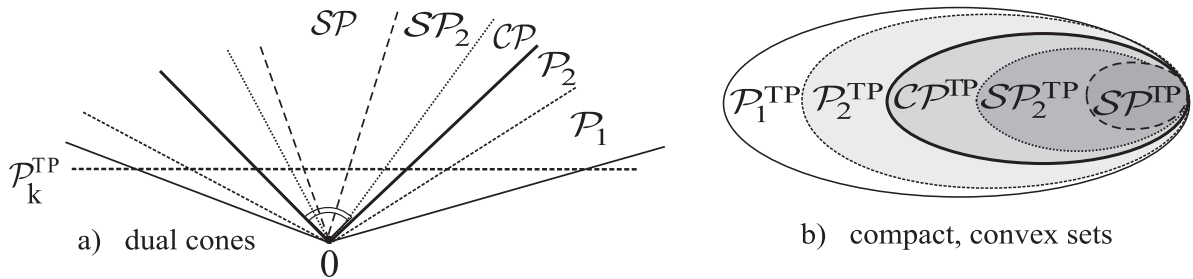}
  \caption{Sketch of sets of maps for the case $d=3$.
    a) Convex cone ${\cal P}={\cal P}_1 $ of positive maps 
   includes the cone ${\cal P}_2 $ of $2$--positive maps
   and its subcone ${\cal P}_3 $ containing the $3$--positive maps,
 which in the case of $d=3$ coincides with the set
$\cal{CP}$ of completely positive maps. It includes the dual sets
of superpositive maps ${\cal SP}_2$ and ${\cal SP}={\cal SP}_1$.
b) Cross-section of the cones with the hyperplane corresponding to the 
trace preserving condition yields a sequence of nested convex bodies,
${\cal P}_k^{TP}$, the volumes of which we aim to estimate.
}
\end{figure}
\vskip 0.3cm 

If we are interested in quantum {\em states}, or density matrices, 
we impose the normalization $\tr \, \rho=1$. 
In other words, we are then 
investigating the base of the corresponding cone, which we will denote by a superscript ``$^1$''.
Thus, in particular,  
\begin{equation} \label{kEnt1}
{\rm Ent}_k^1(\mathbb{C}^d\otimes \mathbb{C}^d) = {\rm Ent}_k(\mathbb{C}^d\otimes \mathbb{C}^d) \cap 
\{M \in \mathcal{B}({\mathbb{C}^d\otimes \mathbb{C}^d}) : \tr \,M = 1 \}
\end{equation}
is the set of 
{\sl $k$-entangled states}
 on $\mathbb{C}^d\otimes \mathbb{C}^d$.
These will be the primary objects of our analysis.
One similarly defines the dual object $ \mathcal{BP}_k^1(\mathbb{C}^d\otimes \mathbb{C}^d)$ of normalized $k$-block positive operators,
which are not necessarily states as they may be non-positive-semi-definite.

On the other hand, when we are interested in quantum {\em maps},  
the {\em trace preserving} constraint  ``$\tr\,\Phi(\rho)={\rm tr}\rho$ for $\rho \in \mathcal{M}_d$''  (or, dually, the {\em unital} constraint)
is more appropriate.  This will be indicated by  a superscript ``$^{TP}$'', for example 
\begin{equation} \label{Pk}
\mathcal{P}_k^{TP} (\mathcal{M}_d) = 
 \mathcal{P}_k \cap  \left \{\Phi: {\calm}_d \to {\calm}_d\   :  
 \  \forall \; \rho \in {\calm}_d \ \tr\,\Phi(\rho)={\rm tr}\rho \right\}
\end{equation}
is the (convex) set of $k$-positive, trace preserving maps and similarly for $\mathcal{SP}_k^{TP} (\mathcal{M}_d) $. 
Note that, under the identifications indicated above,  
\begin{equation} \label{scale}
\mathcal{SP}_k^{TP} (\mathcal{M}_d) \subset d \; {\rm Ent}_k^1(\mathbb{C}^d\otimes \mathbb{C}^d)\ \ \ 
{\rm and } \ \ \  \mathcal{P}_k^{TP} (\mathcal{M}_d) \subset d \; \mathcal{BP}_k^1(\mathbb{C}^d\otimes \mathbb{C}^d).
\end{equation} 
The inclusions are proper for $d>1$ since the unit trace condition involves just one scalar constraint while 
the trace preserving condition leads to $d^2$ independent scalar constraints.
%
%
%

\par \smallskip
We now focus our attention on the set ${\rm Ent}_k^1(\mathbb{C}^d\otimes \mathbb{C}^d)$ of $k$-entangled states.
From the definitions  (\ref{kEnt}) and (\ref{kEnt1}) one concludes that it is  the convex hull of 
the set of {\em pure} $k$-entangled states
\begin{equation} \label{kEntConv}
{\rm Ent}_k^P(\mathbb{C}^d\otimes \mathbb{C}^d) =:   
\big\{|\xi \ket \bra \xi | \, : \, 
\xi \ \hbox{ is } \ k\hbox{-entangled}, |\xi| = 1 \} .
\end{equation}
In other words, this set consists of projections  onto $1$-dimensional subspaces of $\mathbb{C}^d\otimes \mathbb{C}^d$ 
spanned by $k$-entangled vectors.
 It will be also convenient to assign a symbol to the set of {\em vectors} appearing in 
\eqref{kEntConv}. We set
\begin{equation} \nonumber \label{kEntV}
{\rm Ent}_k^V(\mathbb{C}^d\otimes \mathbb{C}^d)= \Big\{ 
\xi = \sum_{j=1}^k  u_j \otimes v_j \; : \;u_j, v_j  \in \mathbb{C}^d \  {\rm for } \  j=1, \ldots , k , 
|\xi|=1 \Big\} \ .
\end{equation}
Note that the sets ${\rm Ent}_k^P$ and ${\rm Ent}_k^V$ ``live'' in different spaces: while 
the former is a subset of ${\rm Ent}_k^1(\mathbb{C}^d\otimes \mathbb{C}^d) \subset \calb(\mathbb{C}^d\otimes \mathbb{C}^d))$,  the latter is a subset of the sphere of  $\mathbb{C}^d\otimes \mathbb{C}^d$. 
We will elaborate on this difference in section \ref{sec:volrad4}.

The tensor product $\C^d\otimes \C^d$ can be canonically identified with the space of 
$d \times d$ matrices $\calm_d$, or with the space of operators on $\C^d$, via the map induced by 
\begin{equation} \label{TenMat}
u \otimes v \to |u\ket\bra v | 
\end{equation}   
(To be precise, operators on $\C^d$ correspond {\em canonically} to the tensor product 
$\C^d\otimes \overline{\C^d}$, but we will not dwell on this distinction.) 
 Under this identification, the set ${\rm Ent}_k^V$ 
corresponds to the set of operators on $\C^d$  whose rank is at most $k$, normalized 
in the Hilbert-Schmidt norm. 
In the sequel we will tend to not  distinguish carefully between these sets, 
and between tensors and operators. 

Since every $\xi \in {\rm Ent}_k^V$ admits a Schmidt decomposition, we also have 
$$
{\rm Ent}_k^V(\mathbb{C}^d\otimes \mathbb{C}^d)= \Big\{ 
\xi \ : \  \xi= \sum_{j=1}^k  s_j \, u_j \otimes v_j \Big\},
$$
where $(u_j)$ and $(v_j)$ are orthonormal sequences in  $\mathbb{C}^d$ and $s_j \geq 0$ with 
$\sum_{j=1}^k s_j^2 =1$.

\vskip 3mm
Our purpose is to  give two-sided estimates for various geometric parameters 
(such as the volume radius or the mean width) 
of  the convex sets 
 ${\rm Ent}_k^1(\mathbb{C}^d\otimes \mathbb{C}^d)$, 
 $\mathcal{P}_k^{TP}(\mathcal{M}_d)$ and $\mathcal{SP}_k^{TP}(\mathcal{M}_d)$. 
Our approach will be to study first the set 
${\rm Ent}_k^V(\mathbb{C}^d\otimes \mathbb{C}^d)$,  
and then to deduce the needed estimates for the remaining sets.

\section{Duality relations and generalities on volume radii}
\label{sec:duality}

In this section we recall the duality relations between the sets
of $k$--positive maps and $k$--superpositive maps.
Due to the classical Urysohn and Santal{\'o} inequalities, and to the relatively 
more recent inverse Santal{\'o} inequality,  duality
allows us to relate estimates for the volumes of the convex bodies
of appropriately normalized trace preserving $k$--positive maps, $k$--superpositive maps
and the corresponding sets of $k$--entangled states.


\subsection{Dual cones and dual sets} \label{sec:dual}

Let  $\mathcal{C}$ be a (closed convex) cone in a real inner product space $\calk$ and  $\mathcal{C}^*$ the dual cone,  
i.e., 
$$\mathcal{C}^* = \{y \in \calk \, : \, \bra y , x\ket \geq 0 \ \hbox{ for all } \ x \in \calc\}.$$ 
As was already indicated above, we have the following duality relation for our cones of interest
\begin{equation}\label{dualstates}
\mathcal{BP}_k(\mathbb{C}^d\otimes \mathbb{C}^d) = \Big({\rm Ent}_k(\mathbb{C}^d\otimes \mathbb{C}^d)\Big)^*,
\end{equation}
where the ambient inner product space is the Hermitian part of $\calb(\mathbb{C}^d\otimes \mathbb{C}^d)$ 
endowed with the Hilbert-Schmidt scalar product $\bra A , B\ket_{HS} := \tr AB^\dagger$ (which in the present context is just $\tr AB$ since $B$ is Hermitian), and similarly 
\begin{equation} \label{dualmaps}
\mathcal{SP}_k(\mathcal{M}_d)  = \Big( \mathcal{P}_k(\mathcal{M}_d) \Big)^*,
\end{equation}
where the duality  for maps  is defined via their Choi matrices \cite{Choi, BZ06} by
$$
(\Phi,\Psi) := \bra C_\Phi, C_\Psi \ket_{HS} .
$$
Since  for closed convex cones we have the bipolar theorem $\big(\calc^*\big)^* = \calc$, 
 the roles of the cones in \eqref{dualstates},  \eqref{dualmaps} can be exchanged.

It is elementary, but not very well known that the duality of cones passes to duality 
$^\circ$ (polarity) of {\em bases}  of cones. Here $^\circ$ is the standard
polar defined by $K^\circ = \{x : \bra x, y \ket~\leq~1 \mbox{ for all } y \in K \}$.  
(In particular, if $K$ is the unit ball in some norm, $K^\circ$ is the unit ball in the dual norm.) 
Let $ \mathcal {C} \subset \calk$ be a closed convex cone and  let  
$e \in \calc \cap \calc^*$ be a unit vector. 
Put  $V^{\rm b} := \{x \in \calk \; : \; \bra x,e \ket =1\}$ and let  
$\mathcal{C}^{\rm b}=\mathcal{C}
\cap V^{\rm b}$ be the base of the cone $\mathcal{C}$. 
Making use of Lemma 1 from \cite{SWZ} one obtains a relation
\begin{equation}
\label{bases}
(\calc^*)^{\rm b} := \calc^* \cap V^{\rm b} = \{y \in V^{\rm b} \; : \; \forall
x \in \calc^{\rm b} \ \ \bra -(y-e),x-e \ket \leq1\} .
\end{equation}
If we think of  $V^{\rm b}$ as a vector space with the origin
at $e$, and of $\calc^{\rm b}$ and  $(\calc^*)^{\rm b}$  as subsets of that
vector space, then  $(\calc^*)^{\rm b} = -(\calc^{\rm b})^\circ$. 
\vskip 3mm
In our case the two dual objects (modulo the reflection with respect to $e$) are 
the appropriately {\em rescaled} sets $d\, {\rm Ent}_k^1(\mathbb{C}^d\otimes \mathbb{C}^d)$
of $k$--entangled states and of 
$k$--positive operators 
$d\, \mathcal{BP}_k^1(\mathbb{C}^d\otimes \mathbb{C}^d)$. 
The rescaling by the factor $d$ is needed since the maximally mixed state $\rho_*=d^{-2}I_{\mathbb{C}^d\otimes \mathbb{C}^d}$ is not of unit length in the Hilbert-Schmidt norm, and the correct   normalization is $e = d^{-1}I_{\mathbb{C}^d\otimes \mathbb{C}^d}=d\,\rho_*$. In the case of maps it is not necessary to 
renormalize; however, since we normally insist on the trace preserving condition  $^{TP}$ 
(which also ``destroys'' precise duality), estimating the size of the sets 
$\mathcal{SP}_k^{TP}(\mathcal{M}_d)$ and  $\mathcal{P}_k^{TP}(\mathcal{M}_d)$ 
requires an additional elementary technical tool (Proposition 6 of \cite{SWZ}).

\subsection {On volume radii, mean widths,  and volumes of dual sets} 
\label{sec:volraddu}


\par
\noindent
Let $K$ be a compact subset of $\mathbb{R}^n$. The volume radius of $K$ is defined as 
$$
{\rm vrad}(K)= \left(\frac{\vol(K)}{\vol (B_2^n)}\right)^{1/n}, 
$$
where $B_2^n$ is the unit Euclidean ball. In other words, 
${\rm vrad}(K)$ is the radius of a Euclidean ball, whose volume is equal to that of $K$.  

Another measure of the size of $K$ is the mean width defined by
$$
w(K) = 2  \int_{S^{n-1}} h_K(u) \, du,
$$
where $du$ is the normalized measure on $S^{n-1}$ and $h_K(u) = \max_{x \in K} \langle x, u \rangle $ 
is the support function of $K$. 
Urysohn's inequality (see, e.g.,   \cite{pisier} or  \cite{Schneider}) 
asserts then that 
\begin{equation} \label{urysohn}
\vrad(K) \ \leq  \ \frac{1}{2}\,  w(K) \; .  
\end{equation}
Urysohn's inequality is usually stated for convex bodies 
(i.e., convex compact subsets of $\mathbb{R}^n$ with nonempty interior), 
but since clearly the width of a set and of its convex hull coincide, 
it is a posteriori true also for non-convex sets.

It is convenient to note that the spherical integral implicit in the definition of the width 
can be expressed as an integral with respect to $\mu_n$, 
the standard Gaussian measure on $\R^n$, 
\begin{equation} \label{gaussian}
\frac{1}{2}\,  w(K) 
= \int_{S^{n-1}} \max_{x \in K} \langle x, u \rangle du 
= \gamma_n\  \int_{\R^n} \max_{x \in K} \langle x, y \rangle \, d\mu_n(y) ,
\end{equation}
where $\gamma_n = \frac{\Gamma(n/2)}{\sqrt{2}\Gamma(n/2+1/2)} \sim \frac 1{\sqrt{n}}$. 
In turn, the Gaussian integral can be interpreted as the expected value of the maximum of 
a Gaussian process. Such quantities have been extensively studied in probability theory. 
In particular, the Dudley's inequality (\cite{dudley}, or see \cite{pisier},  Theorem 5.6) asserts that 
\begin{equation} \label{Dudley}
 \int_{\R^n} \max_{x \in K} \langle x, y \rangle \, d\mu_n(y)
\ \leq \ C \int_0 ^{\infty} \sqrt{\log N(K,\varepsilon)} d \varepsilon,
\end{equation}
where  $C>0$ is a universal numerical constant and $N(K, \varepsilon)$ 
(the covering number)  is the smallest number $N$ such that there are points
 $x_1 \dots, x_N$ such $K \subset \cup_{i=1}^N x_i +\varepsilon B_2^n$ 
 (or,  more generally, in an arbitrary metric space, 
 the smallest number of balls of radius $\ep$ whose union covers~$K$). 
 The expression on the right hand side of \eqref{Dudley} is sometimes called 
 {\em the entropy integral}.

\subsection{Santal\'o and inverse Santal\'o inequalities} \label{sec:santalo}

The classical Santal\'o inequality  \cite{Sa} asserts that 
if $K \subset \mathbb{R}^m$ is a $0$-symmetric convex body and $K^\circ$ its
polar body, then
${\vol (K)} \;{\vol (K^\circ)} \ \le \ \big({\vol \big(B_2^m\big)}\big)^2$ or,
in other words, 
\begin{equation}\label{santalo}
{\rm vrad (K)}\; {\rm vrad (K^\circ)} \leq 1 .
\end{equation}
Moreover, the inequality holds also for 
not-necessarily-symmetric convex sets after an appropriate translation.
In particular,  if the origin is the centroid of $K$ or of $K^\circ$, a condition
that will be satisfied for all sets we will consider in what follows.
Even more interestingly, there is a converse inequality  \cite{BM}, usually 
called ``the inverse Santal\'o inequality,"
\begin{equation}\label{revsantalo}
{\rm vrad (K)}\; {\rm vrad (K^\circ)} \geq c  
\end{equation}
for some universal numerical constant $c>0$, independent of the convex body $K$ 
(symmetric or not) and, most notably, of 
its dimension $m$. An argument yielding 
reasonable value of $c$, particularly  for {\em symmetric} bodies,  was 
given by  Kuperberg \cite{kuperberg}.

The inequalities (\ref{santalo}) and (\ref{revsantalo}) together imply that, 
under some natural hypotheses (which are verified in most 
of cases of interest), the volume radii of a convex body and of its polar are 
approximately (i.e., up to a multiplicative universal numerical constant) 
reciprocal. 

Whenever an upper bound on volume radius is obtained via an estimate 
for the mean width (and then applying \eqref{urysohn}), a lower 
bound on the size of the polar body can be derived without resorting to 
inverse Santal\'o inequality. Instead, one may use the following 
elementary fact
\begin{equation} \label{invurysohn}
{\rm vrad} \big(K^\circ\big) \geq \frac 12 \Big(w(K)\Big)^{-1},
\end{equation}
which is just a consequence of H\"older inequality -- see e.g. 
Appendix A in \cite{s4}.

\section{Volume radii for the set of $k$--entangled states}
\label{sec:volrad4}

\par\noindent

We are now in a position to derive the key results of this work:
bounds for the volume radius of 
${\rm Ent}_k^1(\mathbb{C}^d\otimes \mathbb{C}^d)$,   the set 
of $k$--entangled states of a $d \times d$ system.
We start by estimating the covering numbers of ${\rm Ent}_k^1(\mathbb{C}^d\otimes \mathbb{C}^d)$. 
Then we will use the inequality  (\ref{Dudley}) 
 and the identity (\ref{gaussian}) to estimate the 
mean width $w\big({\rm Ent}_k^1(\mathbb{C}^d\otimes \mathbb{C}^d)\big)$, and subsequently 
the inequality  (\ref{urysohn})  to obtain an upper bound on its volume radius.  A lower 
bound on the volume radius will follow from inequalities for some norms associated   
with the sets ${\rm Ent}_k^1(\mathbb{C}^d\otimes \mathbb{C}^d)$.
Due to the Choi-Jamio{\l}kowski isomorphism one deduces then  
analogous estimates for the volumes of the
set $\mathcal{SP}_k^{TP}(\mathcal{M}_d)$ of $k$--superpositive maps.
Eventually, by duality relations, 
these results lead to bounds for the volume radius of the
set $\mathcal{P}_k^{TP}(\mathcal{M}_d)$ of trace preserving 
$k$--positive maps acting on the $d$-dimensional system.

Before proceeding, we offer a few comments on the relation between 
the unit vectors $\xi \in S_{\C^n} = \{\xi \in \C^n: |\xi|=1\}$,  the cosets  $[\xi] \in S_{\C^n}/S_{\C^1} = \C P^{n-1}$
 (the complex projective space), and the rank one projections $|\xi \ket \bra \xi | \in \calm_n$.
 These distinct objects usually are not carefully distinguished in the physics literature 
 since,  most of the time, no confusion arises. However, here we need to be careful: 
 even though the correspondence 
  $[\xi] \to |\xi \ket \bra \xi |$ is a bijection between $S_{\C^n}/S_{\C^1}$ and the set of 
  pure states on $\C^n$,  these two distinct ``identities'' lead 
to different metric structures. 
For the projective space ${\mathbbm C}P^{n-1}=
S_{\C^n}/S_{\C^1}$, the canonical metric is induced by 
the Euclidean distance on $S_{\C^n} \subset \C^n$ and  
the quotient map  $ \xi \to [\xi]$,
i.e., the distance between $[\xi]$ and $[\eta]$ is 
$\min_{z\in \C, |z|=1} |\xi - z\eta| = 2^{1/2}(1-|\bra \xi|\eta\ket|)^{1/2}$.
This distance, based on the Bures metric (see \cite{BZ06}),
differs from the Hilbert-Schmidt metric,
natural to measure the distance between two density operators,
 $\|  |\xi \ket \bra \xi | -  |\eta \ket \bra \eta |\|_{HS}
   = 2^{1/2}(1-|\bra \xi|\eta\ket|^2)^{1/2}$. 
The ratio between the Hilbert-Schmidt and the Bures distance 
{\em for pure states} is $(1+|\bra \xi|\eta\ket|)^{1/2}$,
which can take any value between $1$ and $\sqrt{2}$.
This is actually good news since it tells us that when passing 
from one framework to the other we {\em at worst}  need  to pay the price of a 
dimension independent factor of $\sqrt{2}$.

\subsection{Upper bound for
     $\vrad\left({\rm Ent}_k^1(\mathbb{C}^d\otimes \mathbb{C}^d)\right)$}

Since, as mentioned earlier,  the width of a set is the same as that of its convex hull,
we will be estimating via the Dudley's inequality \eqref{Dudley}
 the mean width of 
${\rm Ent}_k^P(\mathbb{C}^d\otimes \mathbb{C}^d)$, 
 the set of the 
extreme points of the set  ${\rm Ent}_k^1(\mathbb{C}^d\otimes \mathbb{C}^d)$ of 
$k$-entangled states.
In turn, by the remark above, the latter problem reduces -- at the cost of a multiplicative 
factor  not exceeding $\sqrt{2}$ -- to estimating the entropy 
integral of the sets of $k$-entangled {\em vectors}  
${\rm Ent}_k^V(\mathbb{C}^d\otimes \mathbb{C}^d) \subset S_{\mathbb{C}^d\otimes \mathbb{C}^d}$. 
We will employ the identification indicated in  \eqref{TenMat}, 
i.e.,  
\begin{equation}\nonumber \label{EntInMd}
 {\rm Ent}_k^V(\mathbb{C}^d\otimes \mathbb{C}^d) \sim
 \Big\{ 
\sum_{j=1}^k s_j |u_j \ket \bra  v_j|  \; : \; s_j \geq 0, \; 
\sum_{j=1}^k s_j^2 =1\Big\} \subset \calm_d,
\end{equation}
where $(u_j)$ and $(v_j)$ vary over  orthonormal sequences in  $\mathbb{C}^d$, 
and the 
problem reduces to finding, for $\ep \in (0,1)$,   $\varepsilon$-nets of this set 
of $d \times d$ matrices  (with respect to the 
Euclidean, or Hilbert-Schmidt metric)  with good bounds on their cardinalities.

Given  
$\tau =\sum_{j=1}^k t_j |u_j \ket \bra  v_j| \in  {\rm Ent}_k^V(\mathbb{C}^d\otimes \mathbb{C}^d)$, set
\begin{equation}\label{EF}
E=E_\tau=\mbox{span}\{u_i: 1 \leq i \leq k\},  \hskip 5mm  F=F_\tau=\mbox{span}\{v_i: 1 \leq i \leq k\} 
\end{equation}
\par
\noindent
and let $T$ be the matrix $\tau$ considered as an operator acting from $F$ to $E$. 
Thus,  to each $\tau \in {\rm Ent}_k^V(\mathbb{C}^d\otimes \mathbb{C}^d)$ there correspond $(E,F) \in G_{d,k}\times G_{d,k}$ 
and $T \in S_{HS(F,E)}$ such that $\tau= T P_F$, where $P_H$  stands for the orthogonal projection of $ \mathbb{C}^d$ onto $H$.  Accordingly, the problem reduces to 
estimating  the appropriate
covering numbers  of  the Grassmann manifold $G_{d,k}$ and of $S_{HS(F,E)}$, 
which -- geometrically -- is just the unit sphere in a 
$k^2$-dimensional (complex) Euclidean space.  

Before proceeding, we shall make more precise the metric structure of  $G_{d,k}$ implicit in the concept of a net needed for our construction. It will be based on the operator norm  $\|\cdot\|_{op}$ on $\calm_d$; if $E, E'$ are $k$-dimensional subspaces of $\C^d$, we set 
$$
d_{op}(E, E') :=  \min \{ \|U-I\|_{op} \; : \;  U\in \mathcal{U}(d), UE=E'\},
$$
where $ \mathcal{U}(d)$ is the unitary group.
We note that $d_{op}$ is the quotient distance induced by the {\em extrinsic} 
operator norm on $\calm_d \supset  \mathcal{U}(d)$ and {\it not } an intrinsic (geodesic) distance.
The corresponding 
intrinsic distance is the largest principal angle between $E$ and $E'$ and is 
induced in the same way by the geodesic distance on $\mathcal{U}(d)$. 
For clarity, we note that if $\alpha$ is the 
largest principal angle between $E$ and $E'$, then $d_{op}(E, E') = |e^{i\alpha} - 1| =2 \sin (\alpha/2)\leq \alpha$, and it is elementary to check that then $\|P_E - P_{E'}\|_{op} = \sin \alpha \leq d_{op}(E, E') $ (this is not going to be used). 
We now claim that an $\ep$-net $\caln_1$ on  $G_{d,k}$ (with respect to $d_{op}$) and $\ep$-nets 
(with respect to the Hilbert-Schmidt  norm $\|\cdot\|_{HS}$) 
$\caln_{E,F}$ on $ S_{HS(F,E)} \sim S_{\C^{k^2}}$ for  $E,F \in \caln_1$ lead to a $3\ep$-net  
$$\caln = \{T P_F \; : \; E,F \in \caln_1,  T  \in \caln_{E,F}\}$$
in ${\rm Ent}_k^V(\mathbb{C}^d\otimes \mathbb{C}^d)$. Indeed,  consider an arbitrary element of ${\rm Ent}_k^V(\mathbb{C}^d\otimes \mathbb{C}^d)$ induced by  $E^\prime, F^\prime  \in G_{d,k}$ 
and  $T^\prime: F^\prime \rightarrow E^\prime$, i.e., $T^\prime P_{F^\prime}$.
 Let $E$ and  $F$  in $\caln_1$ be such that 
 \begin{equation*}
 d_{op}(E, E')\leq \ep, \hskip 5mm 
 d_{op}(F, F')\leq \ep.
 \end{equation*}
 Next, let $U, V \in \mathcal{U}( d)$  be such that $VF=F^\prime$,  $UE^\prime=E$ and 
 \begin{equation*}
 \|V-I\|_{op} \leq 
 d_{op}(F, F') \leq  \ep \ \   \text{and} \  \  
  \|U-I\|_{op} \leq  
  d_{op}(E, E')\leq \ep
 \end{equation*}
  Set  $S= U
T^\prime V$. Then 
  $S\in S_{HS(F,E)}$ and consequently $S P_F$ 
  can be approximated  within $\ep$ (in the Hilbert-Schmidt  norm) by an element of $\caln$. 
  Thus it will follow that $\caln$ is an $3\ep$-net of ${\rm Ent}_k^V(\mathbb{C}^d\otimes \mathbb{C}^d)$ 
   if we show that $S P_F$ is within $2\ep$ of $T^\prime P_{F'}$. To that end, we note first that 
  $S P_F = UT^\prime V P_F = UT^\prime P_{F'} V$ (the second equality because $P_{F^\prime}=VP_FV^\dagger$) and so 
  \begin{eqnarray*}
\|S P_F- T^\prime P_{F'}\|_{HS} &=&
\|U T^\prime P_{F'} V- T^\prime P_{F'}\|_{HS}\\
&\leq&  \|U T^\prime P_{F'} V- T^\prime P_{F'} V\|_{HS} + 
\|T^\prime P_{F'} V- T^\prime P_{F'}\|_{HS}\\
&\leq& \|U -I\|_{op}\|T^\prime P_{F'} V\|_{HS} +\|T^\prime P_{F'}\|_{HS} \| V- I\|_{op}\\
&\leq& \ep\|T^\prime \|_{HS}+ \ep\|T^\prime \|_{HS} = 2 \ep
\end{eqnarray*}
as required.
It now remains to collect known estimates on  covering numbers (with respect to the appropriate metrics) of $G_{d,k}$ and $S_{\C^{k^2}}$. 
In both cases these estimates are of the form $N(K,\delta) \leq \big(C/\delta\big)^{\dim_\R K}$, 
where $C>0$ is a  universal constant
(for $G_{d,k}$, see Remark 8.4 in \cite{s1} for the statement and \cite{s2, s3} for proofs and more general setting; the case of $S_{\C^{m}}$ is classical, see \cite{pisier},
Lemma 4.10). 
This yields  an estimate 
$$
N({\rm Ent}_k^V(\mathbb{C}^d\otimes \mathbb{C}^d), \ep) \leq \#\caln \leq\big(3C/\ep\big)^{2k^2} \Big(\big(3C/\ep\big)^{4k(d-k)}\Big)^2 
= \big(C'/\ep\big)^{2k(4d-3k)} \leq \big(C'/\ep\big)^{8kd} 
$$
as we assumed that $k\leq d$.
By prior remarks, passing from ${\rm Ent}_k^V(\mathbb{C}^d\otimes \mathbb{C}^d)$ to ${\rm Ent}_k^P(\mathbb{C}^d\otimes \mathbb{C}^d)$ introduces at worst 
an extra $\sqrt{2}$ factor, or replacing the constant $C'$ by $C_1=\sqrt{2}C'$.
Accordingly,   the integrand  $\sqrt{\log N(K,\varepsilon)} $ in (\ref{Dudley}) 
 for $K={\rm Ent}_k^P(\mathbb{C}^d\otimes \mathbb{C}^d)$ 
is at most $\sqrt{8kd} \, \sqrt{\log {(C_1/\ep)}}$,  while the upper limit of integration 
is  $1$.  This means that the singularity at $\ep = 0$ is integrable, and the final estimate 
is 
\begin{eqnarray}\label{S-above}
\vrad\left({\rm Ent}_k^1(\mathbb{C}^d\otimes \mathbb{C}^d)\right)  
&\leq& \frac 12 w({\rm Ent}_k^1(\mathbb{C}^d\otimes \mathbb{C}^d)) = \frac 12 w({\rm Ent}_k^P(\mathbb{C}^d\otimes \mathbb{C}^d))  \leq C \gamma_{d^4} C' \sqrt{kd}  \nonumber \\
&\leq& C_0\frac {k^{1/2}}{d^{3/2}}
\end{eqnarray}
For $k=1$ (separable states) and for $k=d$ (all states) the above bound is known to give the correct order 
(see, e.g., \cite{AS}).
In the next subsection we are going to show
that the obtained bound is a tight estimate for the volume 
also for the intermediate cases $k=2,\dots,d-1$.



\subsection{The lower bound} 

We start with an inequality concerning norms on $\calh = \C^d\otimes\C^d$. 
Given $k \in \{1, 2, \ldots, d\}$,  
we set, for $\xi \in \C^d\otimes\C^d$
\begin{equation} \label{k-norms}
\|\xi \|^{(k)} := \max _{\, \zeta \in {\rm Ent}_k^V(\mathbb{C}^d\otimes \mathbb{C}^d)} | \bra \xi , \zeta \ket |
\end{equation}
(Under the identification \eqref{TenMat},  $\|\cdot \|^{(k)}$ corresponds to the operator norm and $\|\cdot \|^{(d)}$ to the 
Hilbert-Schmidt/Frobenius norm.) 
We will need the following  elementary inequality 
\begin{equation} \label{auxil}
\|\xi \|^{(k)}  \geq \sqrt{k/d} \;\|\xi \|^{(d)} =  \sqrt{k/d} \; |\xi |\ \ \ {\hbox{ for all }} \ \xi \in \C^d\otimes\C^d .
\end{equation}
The norms defined by \eqref{k-norms} were studied -- independently of this paper -- 
in \cite{JK}, where they were denoted $\|\cdot\|_{s(k)}$. (Note that the arxiv version of \cite{JK} 
corrects some minor errors present in the published version that are relevant to our argument.) 
In particular, \eqref{auxil} is a special case of Corollary 3.4 in  \cite{JK}. We will sketch 
the argument for completeness. 
To this end, let  $\xi = \sum_{j=1}^d s_j \, u_j\otimes v_j$ be the Schmidt 
decomposition and, for $\Lambda \subset  \{1, 2, \ldots, d\}$, set 
$ \xi_\Lambda =  \sum_{j \in \Lambda}s_j \, u_j\otimes v_j$. If $\Lambda$ 
varies over all subsets of $\{1, 2, \ldots, d\}$ of size $k$, then $\E \xi_\Lambda = \frac kd \xi$, 
$\E |\xi_\Lambda|^2 = \frac kd |\xi|^2$ 
(where $\E$ is the average over all choices of $\Lambda$) 
and $\xi_\Lambda/|\xi_\Lambda|  \in 
{\rm Ent}_k^V(\mathbb{C}^d\otimes \mathbb{C}^d)$ for all 
$\Lambda$ such that $\xi_\Lambda \neq 0$. We have, on the one hand, 
$$
\E \bra \xi , \xi_\Lambda \ket = \frac kd \bra \xi ,  \xi \ket = \frac kd |\xi|^2
$$
while, on the other hand, ignoring $\Lambda$'s for which $\xi_\Lambda = 0$, 
\begin{eqnarray*}
\E \bra \xi , \xi_\Lambda \ket &=& 
\E \Big( \Big\bra \xi , \frac{\xi_\Lambda}{|\xi_\Lambda|} \Big\ket  |\xi_\Lambda|  \Big) 
\leq \|\xi\|^{(k)} \E |\xi_\Lambda| \\
&\leq&  \|\xi\|^{(k)} \left(\E |\xi_\Lambda|^2\right)^{1/2} =  \|\xi\|^{(k)} \sqrt{\frac kd} |\xi|
\end{eqnarray*}
and it remains to compare the two expressions.

We now want to show that the symmetrized set of (mixed) $2k$-separable states, 
i.e., ${\rm conv} \big(-{\rm Ent}_{2k}^1(\mathbb{C}^d\otimes \mathbb{C}^d) 
\cup {\rm Ent}_{2k}^1(\mathbb{C}^d\otimes \mathbb{C}^d)\big)$, 
considered as a subset of the $\R$-linear 
subspace of $\calb(\C^d\otimes\C^d)$ consisting of Hermitian matrices, 
contains a Hilbert-Schmidt ball of radius 
$\frac {k^{1/2}}{d^{3/2}}$.  A standard argument based on Rogers-Shephard 
inequality  -- as in \cite{s4}, Appendix C -- will then  imply that 
$ \vrad \bigl( {\rm Ent}_{2k}^1 (\mathbb{C}^d\otimes \mathbb{C}^d)\bigr)  \geq \frac 12  \frac {k^{1/2}}{d^{3/2}}$, hence 
\begin{equation}\label{lower}
\vrad \bigl( {\rm Ent}_{k}^1 (\mathbb{C}^d\otimes \mathbb{C}^d)\big)  \geq \frac 12  \frac {\lfloor k/2 \rfloor ^{1/2}}{d^{3/2}},
\end{equation}
which gives the needed lower bound. (Note that we can assume that $k\geq 2$ since the 
case $k=1$ was handled already in \cite{AS}.)

To show that ${\rm conv} \big(-{\rm Ent}_{2k}^1(\mathbb{C}^d\otimes \mathbb{C}^d) 
\cup {\rm Ent}_{2k}^1(\mathbb{C}^d\otimes \mathbb{C}^d)\big)$  
contains the appropriate Hilbert-Schmidt ball we will argue by duality: 
we will prove that the support function of 
$-{\rm Ent}_{2k}^1(\mathbb{C}^d\otimes \mathbb{C}^d) 
\cup {\rm Ent}_{2k}^1(\mathbb{C}^d\otimes \mathbb{C}^d)$ is 
bounded from below by $\frac {k^{1/2}}{d^{3/2}}$. 
To that end, consider an arbitrary 
Hermitian operator  $A$ on $\C^d\otimes \C^d$ 
with $\|A\|_{HS}=1$. We need to show that
\begin{eqnarray*}
\max_{M\in  {\rm Ent}_{2k}^1(\mathbb{C}^d\otimes \mathbb{C}^d)} |\,{\tr} \, A M\,| &=& \max_{|\eta\ket \in  {\rm Ent}_{2k}^V(\mathbb{C}^d\otimes \mathbb{C}^d)} \big|\,{\tr} \, A \,|\eta\ket\bra \eta |\, \big|\\
&=&\max_{|\eta\ket \in {\rm Ent}_{2k}^V(\mathbb{C}^d\otimes \mathbb{C}^d)} \big| \bra   \eta | A \,|\eta\ket \big| \\
&\geq &\frac {k^{1/2}}{d^{3/2}} \ .
\end{eqnarray*}
The equalities are immediate; 
 the inequality  will be shown by establishing a chain of identities and 
inequalities 
\begin{eqnarray}
\max_{|\eta\ket \in {\rm Ent}_{2k}^V(\mathbb{C}^d\otimes \mathbb{C}^d)} \big| \bra   \eta | A \,|\eta\ket \big| &\geq &
\max_{|\phi\ket , |\psi\ket \in {\rm Ent}_{k}^V(\mathbb{C}^d\otimes \mathbb{C}^d)}
{\rm Re} \big(\bra   \phi | A \,|\psi\ket \big) \nonumber \\
&=&\max_{|\phi\ket , |\psi\ket \in {\rm Ent}_{k}^V(\mathbb{C}^d\otimes \mathbb{C}^d)} 
\big| \bra   \phi | A\,|\psi\ket \big|  \label{chain} \nonumber\\
&\geq &\frac {k^{1/2}}{d^{3/2}} \ . 
\end{eqnarray}
Again, the equality is easy: it  follows from the fact that the quantity 
under the third maximum does not 
change when we multiply $|\phi\rangle $ or $|\psi\rangle $
 by $z \in \C$ with $|z|=1$.

For the second inequality in (\ref{chain}),  
we notice that for a fixed $ |\psi\ket \in \C^d\otimes \C^d$,
$$
\max_{|\phi\ket  \in {\rm Ent}_k^V(\mathbb{C}^d\otimes \mathbb{C}^d)} \big|\bra   \phi | A \,|\psi\ket \big|
=  \| A \,|\psi\ket \|^{(k)} \geq 
\sqrt{\frac kd}\, \big| A \,|\psi\ket \big| 
$$
by (\ref{auxil}). 
Next,
$$
\max_{|\psi\ket \in  {\rm Ent}_k^V(\mathbb{C}^d\otimes \mathbb{C}^d)} \big| A \,|\psi\ket \big|\geq 
\max_{|\psi\ket \in  {\rm Ent}_1^V(\mathbb{C}^d\otimes \mathbb{C}^d)} \big| A \,|\psi\ket \big|
\geq 
 \sqrt{ \big\langle  \bigl | A| \psi \rangle \big|^2 \big\rangle_{\psi} } ,
$$
where the symbol $ \big\langle \cdot \big\rangle_{\psi} $  under the square root 
stands for the average,  taken with respect to the natural
product measure, over  
 $\psi \in  {\rm Ent}_1^V(\mathbb{C}^d\otimes \mathbb{C}^d) = 
\{u \otimes v \, : \, u,v \in S_{\C^d}\} \sim S_{\C^d} \times S_{\C^d}$, 
the set of normalized product pure states.
It remains to check that this average equals  
 $$\int_{S_{\C^d}} \int_{S_{\C^d}}  \big| A \,|u\otimes v\ket \big|^2\, du \, dv
 = \frac 1{d^2} \| A \|_{HS}^2 =\frac 1{d^2}
 $$ 
and combine the estimates.

The first inequality in (\ref{chain}) follows via a standard polarization argument, 
with a small additional twist to obtain an estimate without any additional 
multiplicative constants.  First, since $A$ is Hermitian one has
$$
{\rm Re} \big(\bra   \phi | A \,|\psi\ket \big) = {\rm Re} \big(\tr (A \,|\psi\ket \bra   \phi |)\big) 
= \tr \big( A \, {\rm Re}(|\psi\ket \bra   \phi |)\big) \ . 
$$
Next, we have an elementary identity,
$$
\Big|\frac{\phi+\psi}2\Big\ket\Big\bra \frac{\phi+\psi}2\Big| - \Big|\frac{\phi-\psi}2\Big\ket\Big\bra \frac{\phi-\psi}2\Big| =
{\rm Re} \big(\,|\psi\ket \big\bra   \phi | \big).
$$
Combining the two we obtain
$$
{\rm Re} \big(\bra   \phi | A \,|\psi\ket \big) 
= \bra   \eta_1 | A \,|\eta_1\ket  - \bra   \eta_2 | A \,|\eta_2\ket ,
$$
where $\eta_1= |\frac{\phi+\psi}2\ket$ and $\eta_2= |\frac{\phi-\psi}2\ket$.  
Since the Schmidt rank of $\eta_1$ and $\eta_2$ is at most $2k$, 
it follows that 
${\rm Re} \big(\bra   \phi | A \,|\psi\ket \big) \leq 
\big(|\eta_1|^2 + |\eta_2|^2\big)  
\max_{|\eta\ket \in   {\rm Ent}_{2k}^V(\mathbb{C}^d\otimes \mathbb{C}^d)} \big| \bra   \eta | A \,|\eta\ket \big|$.
On the other hand,  the parallelogram identity yields 
$|\eta_1|^2 + |\eta_2|^2 = |\frac{\phi+\psi}2|^2 + |\frac{\phi-\psi}2|^2 
= \frac{|\phi|^2+|\psi|^2}2 = 1$,
which gives the needed first inequality in (\ref{chain}). 

The inequalities (\ref{chain}) may be of interest in themselves. Let us just mention that, 
in the notation of \cite{JK}, they read
\begin{equation} \label{asInJK}
\max_{|\eta\ket \in {\rm Ent}_{2k}^V(\mathbb{C}^d\otimes \mathbb{C}^d)} \big| \bra   \eta | A \,|\eta\ket \big| \geq 
\|A\|_{S(k)} \geq \frac {k^{1/2}}{d^{3/2}} \|A\|_{HS} 
\end{equation}
for $A \in \calb(\mathbb{C}^d\otimes \mathbb{C}^d)$, with the first inequality requiring additionally $A=A^\dg$.

\subsection{Consequences for the remaining sets}

  Since we determined, in \eqref{S-above} and \eqref{lower}, 
the precise asymptotic order of the volume radius of the set 
${\rm Ent}_k^1(\mathbb{C}^d\otimes \mathbb{C}^d)$ as $\Theta\big((k/d^3)^{1/2}\big)$, 
it follows from \eqref{dualstates},  
\eqref{bases} and the discussion in section \ref{sec:santalo} 
 that, for the dual set, 
$\vrad \big(\mathcal{BP}_k^1(\mathbb{C}^d\otimes \mathbb{C}^d)\big) = \Theta\big((kd)^{-1/2}\big)$ 
(remember the ``rescaling'' issue pointed out in the last paragraph of section \ref{sec:dual}).

To deduce sharp estimates on the 
volume radius of $\mathcal{SP}_k^{TP}(\mathcal{M}_d)$ 
and $\mathcal{P}_k^{TP}(\mathcal{M}_d)$ we appeal to \eqref{scale}
 and then to Proposition 6 of \cite{SWZ}.   Heuristically, Proposition 6 of \cite{SWZ} says that 
 if an  $m$-dimensional convex body $K$ is ``reasonably balanced,'' then all its central 
 sections whose codimension is ``substantially smaller'' than $m$ have volume radii 
 close to that of $K$. In the present setting the dimensions and codimensions are 
 exactly the same as in the applications discussed in \cite{SWZ}, and the bodies we 
 consider are intermediate with respect to those in \cite{SWZ}. Accordingly, all 
 the arguments carry over and the heuristic principle described above can 
 be rigorously shown to hold, and we can conclude that  the volume radii of the 
 bodies on the left hand side of each of the inclusions in \eqref{scale} 
 are essentially the same as those of the respective bodies on the right hand side. 
 In other words,  $\vrad \big(\mathcal{SP}_k^{TP}(\mathcal{M}_d)\big) = \Theta\big((k/d)^{1/2}\big)$ and $\vrad \big(\mathcal{P}_k^{TP}(\mathcal{M}_d)\big) = \Theta\big((d/k)^{1/2}\big)$. 

\medskip

 \section{Concluding Remarks} \label{sec:conclusion}
 
In this work we obtained explicit and asymptotically sharp 
estimates for the volume radius
(in the sense of Hilbert-Schmidt measure) of the convex body 
${\rm Ent}_k^1(\mathbb{C}^d\otimes \mathbb{C}^d)$
of normalized $k$--entangled states.
For $k=1,\dots, d-1$ these bodies form a nested family of subsets 
of the set of all  states on a bipartite $d\times d$ system, 
which in the present notation coincides with 
${\rm Ent}_d^1(\mathbb{C}^d\otimes \mathbb{C}^d) \subset {\cal M}_{d^2}$.

Making use of the Choi-Jamio{\l}kowski isomorphism we deduce then 
 bounds for the volumes of the sets ${\cal SP}_k^{TP}$
of trace-preserving $k$--superpositive maps, also called $k$--entanglement 
breaking channels.
Finally, appealing to the Santal{\'o}  and inverse Santal{\'o} inequalities, 
which relate the volumes of two dual sets, 
we obtain tight  estimates for the volumes of   sets ${\cal P}_k^{TP}$ 
of trace preserving $k$--positive maps,  
acting on density matrices of a given  size $d$. (This is again a nested 
family, with extreme cases $k=1$ and $k= d$  corresponding 
respectively to 
positivity preserving and completely positive quantum maps.)

Our findings show that, in large dimension,  the property of 
``additional degrees of positivity'' is uncommon.  On the other hand, allowing 
``additional degrees of entanglement''  is a major relaxation.  
While our methods allow only to compare the set of  $k$--entangled states  
with that of $ak$--entangled states, where $a>1$ in a universal constant 
(and similarly for  maps), the estimates obtained 
strongly suggest that, for large dimension $d$,
the set of $k$--entangled states covers only
a small fraction of the larger set of $(k+1)$--entangled states,
with the trend particularly pronounced for small $k$'s. 
Indeed, we did show that, for $\calh=\mathbb{C}^d\otimes \mathbb{C}^d$, the ratio 
$R_{k,d}:=\frac{\vrad \big({\rm Ent}_k^1(\calh)\big)}{ (k/d^3)^{1/2}}$ 
verifies $c_0\leq R_{k,d} \leq C_0$ for some positive constants $c_0, C_0$ independent of 
$k,d$. If, instead, we had $R_{k,d} = c(d)$, with $c(d) \in [c_0, C_0]$ independent of $k$, 
it would follow that  
$\frac{\vol\big({\rm Ent}_{k+1}^1(\calh)\big)}{\vol\big({\rm Ent}_k^1(\calh)\big)} = \Big(1+\frac 1 k\Big)^{(d^2-1)/2}$.   While it is hard to expect that the 
equality $R_{k,d} = c(d)$ holds {\sl precisely}, it would hold {\em approximately} if the dependence of  
$R_{k,d}$ on the parameters $k,d$ was regular enough (which we can not prove rigorously).  
However,  it {\em does} follow form our estimates that ``in the mean'' the ratios of the volumes 
of successive sets do behave as indicated above, i.e., 
are exponential in $d^2$ for small $k$, and then taper off to  exponential in $d$ 
when $k$ is of order $d$. Further, it follows that if a state is randomly chosen (with respect 
to the Hilbert-Schmidt volume), then the probability that it is $k$ entangled can be 
upper-bounded by    $\big(\frac{Ck}d\big)^{(d^2-1)/2}$ and lower-bounded by     $\big(\frac{ck}d\big)^{(d^2-1)/2}$, where 
$C\geq c>0$ are universal constants.
Similar comments can be made about 
volumes of the sets ${\cal P}_k^{TP}$ of $k$-positive maps, except that in 
that setting the volumes {\em decrease} with $k$.

Of  special importance are the cases involving $k=2$. This is because 
understanding the set of $2$--entangled states is crucial (for example)
for problems related to the distillation of quantum entanglement.
It thus would be of substantial interest to rigorously show that, 
for large dimension $d$,  the ratios $\frac{\vol\big({\rm Ent}_{2}^1(\calh)\big)}{\vol\big({\rm Ent}_1^1(\calh)\big)}$ and $\frac{\vol\big({\rm Ent}_3^1(\calh)\big)}{\vol\big({\rm Ent}_2^1(\calh)\big)}$ 
do indeed behave as predicted by our ``global'' estimates, i.e., that  the set of $2$--entangled states is 
exponentially (in $d^2$) small in comparison to the set of $3$-entangled states, but 
exponentially large with respect to the set of separable states.

\vskip.5cm \noindent 
{\em Acknowledgements}:
It is a pleasure to thank to the Oberwolfach Institute for Mathematics,
where this project was initiated during the December 2009 workshop 
``Geometry of Quantum Entanglement.''
The first-named and the second-named authors were partially 
supported by grants from the National Science Foundation (U.S.A.).
The third-named author acknowledges support under the grant
 N N202 090239 of Polish Ministry of Science and Higher Education,
 Foundation for Polish Science and European Regional Development Fund
(agreement no MPD/2009/6),
and under the SFB Transregio-12 project financed by the DFG.


\end{document}